\documentclass[pra, a4paper, twocolumn]{revtex4}   
\usepackage{graphicx}

\begin{document}

\title {All optical sensor for automated magnetometry based on
  coherent population trapping}

\author{ J.Belfi, G.Bevilacqua, V.Biancalana, Y.Dancheva, L.Moi}

\address{CNISM-Unit\`a di Siena, Dipartimento di Fisica -
Universit\`a di Siena, via Roma
  56, 53100 Siena, Italy}

\begin{abstract}
An automated magnetometer suitable for long lasting
measurement under stable and controllable experimental conditions
has been implemented. The device is based on Coherent Population
Trapping (CPT) produced by a multi-frequency excitation. CPT 
resonance is observed when a frequency comb,
generated by diode laser current modulation, excites Cs atoms
confined in a $\pi/4\times(2.5)^2\times1~\textrm{cm}^3$, 2 Torr
$N_2$ buffered cell.  A fully
optical sensor is connected through  an optical fiber to the laser
head allowing for truly remote sensing and minimization of the field
perturbation. A detailed analysis of the CPT resonance
parameters as a function of the optical detuning has been made in
order to get high sensitivity measurements. The
magnetic field monitoring performances and the best sensitivity
obtained in a balanced differential configuration of the  sensor
are presented.

OCIS {120.4640, 020.1670, 300.6380.}

This work has been submitted to JOSA B for publication

[Josa B (7) 2007]

\end{abstract}

\maketitle

\section{Introduction}
Atomic magnetometers, developed since 1960's \cite{Blo62}, have
today a central role in the field of high sensitive magnetometry
with important applications in geophysics, medicine, biology,
testing of materials and of fundamental physics symmetries.
Technical advances reached in the last years, will make optical
magnetometers more suitable than SQUIDs in most of these
applications because of the comparable - in some cases, even
better \cite{Kom03} - sensitivity and the possibility to operate at
room temperature with no need of cryogenic cooling.

Recently a direct measure of geophysical-scale field, with the
impressive sensitivity of $60~\mbox {fT}/\sqrt{\mbox {Hz}}$ was
demonstrated in a Non-linear Magneto-Optical Rotation experiment
\cite{Aco06}.

Detection of weak biological-scale magnetic fields
(magneto-cardiometry) has been demonstrated using optically pumped
magnetometers \cite{Wei03a, Wei03b} based on the `double
optical-RF excitation'. Direct RF magnetic excitation, anyway, does not
permit the realization of fully optical sensors as RF coils have to be placed
next to the vapor cell.

All-optical sensors based on Coherent Population Trapping (CPT)
effect can be instead built. CPT effect occurs when two long-lived
ground states are  coupled to a common excited state by two
coherent laser fields. When the frequency difference of the laser
fields exactly matches the frequency separation of two not coupled
ground levels, the population is trapped in the so called `dark'
state. This  is a quantum-mechanical superposition of both ground
states that is not coupled to the laser fields. An accumulation of
population in such coherent state gives rise to a resonant
transparency \cite{Alz76, Ari76}. CPT resonances have
line-widths much narrower than the natural line-width of
the corresponding optical transitions. This makes them particularly suitable
for precision spectroscopy applications
in many other fields besides magnetometry \cite{Scu92, Fle94} like
metrology \cite{Tho82, Hem93}, detection of gravitational waves
\cite{Cav81}, laser physics \cite{Har89, Koc92} and laser cooling\cite{Coh90}.

In this work we present the characteristics of an all-optical
magnetometer, working as a compact, automated device able to
measure magnetic fields in wide range of amplitudes and time-scales,
for example the daily Earth magnetic field variations or 
weak signals varying in the msec time-scales, superposed on the Earth
magnetic field. The sensor works in the magnetically polluted  environmental 
conditions typical of a scientific laboratory. Neither additional RF
magnetic field excitation nor particular magnetic field shielding are
necessary. 
 
The principle of operation of our CPT magnetometer can be
described as follows. Couples of laser fields, such that their
frequency separation matches the energy splitting between the
Zeeman sub-levels of a given hyperfine ground state of Cs, are
produced by frequency modulation  of the diode laser junction
current in the $10$~kHz~-~$10$~MHz range. The obtained frequency
modulated radiation is characterized by a very high modulation
index ($\sim$ $10^3$) and the spectral structure of the laser can
be seen as a comb of coherent modes with an overall width of the
order of the Doppler broadened optical transition. With such a
broad band excitation almost all the atomic velocity classes
interact  with resonant light and furthermore the power absorbed
by each single class is very low. The first feature allows us to
increase the resonance contrast, the second one reduces the power
broadening. It is worth noting that this solution allows us to work
without  complex laser phase locking systems and, furthermore, without
expensive and bulky highly stable microwave oscillators that are, instead,
necessary in the case of CPT generation on Zeeman sub-levels of
different hyperfine ground states\cite{Aff02}.

\section{Experimental setup and measurement procedure}
\begin{figure}[htbp]
  \centering
  \includegraphics[width=7cm]{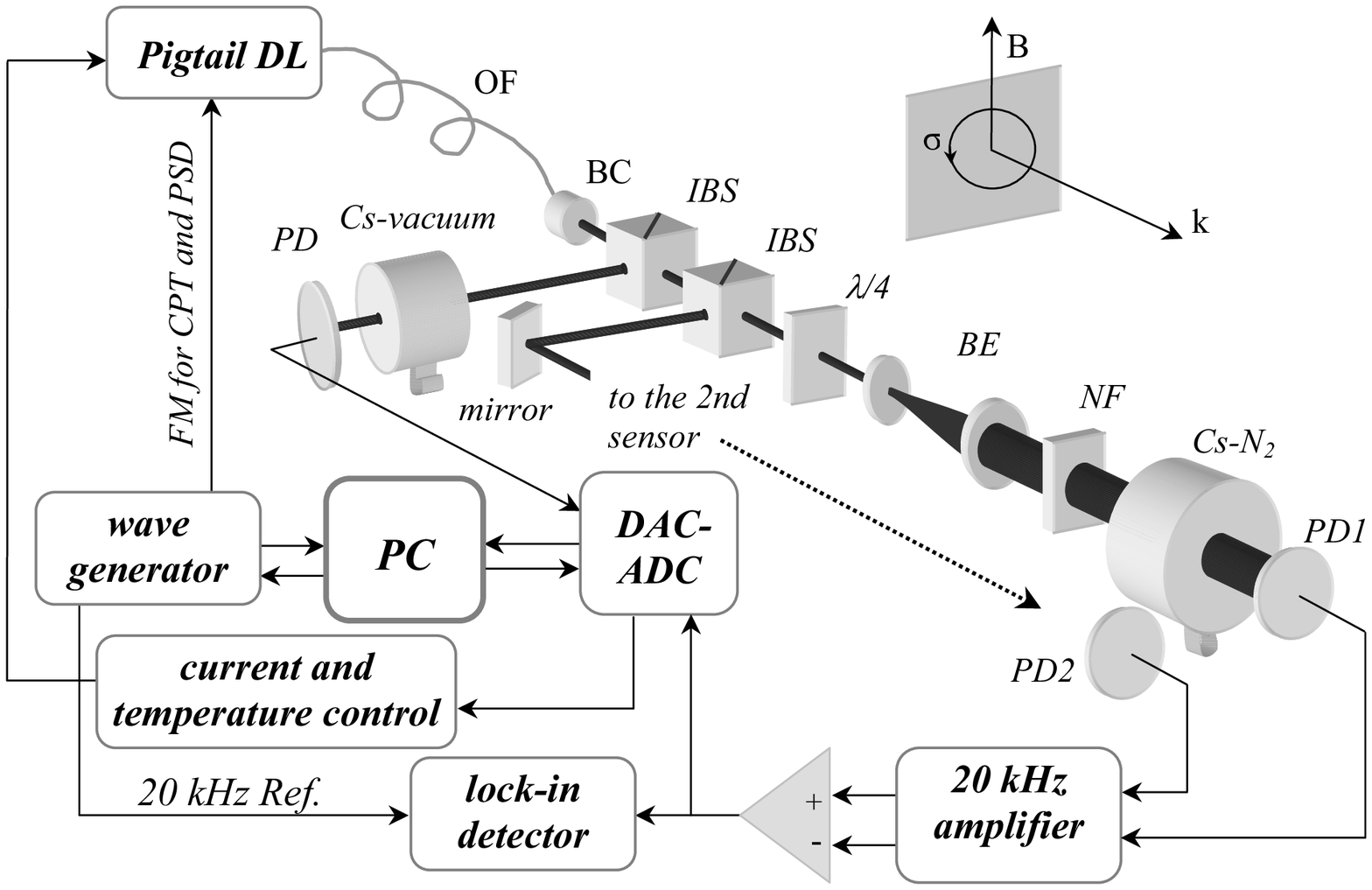}
    \caption{Experimental setup. OF: optical fiber, BC: beam
    collimator, IBS: intensity beam splitter, BE: beam expander, NF:
    neutral filters, PD: photo diode.  \label{setup}} 
\end{figure}

A sketch of the experimental setup is presented in
Fig.\ref{setup}. The sensor measures the resonant absorption of laser
light by Cs vapor contained in a cylindrical cell 2.5~cm in diameter and
1~cm in length. The cell is kept at room temperature and contains
2~Torr of $\mbox N_2$ as a buffer gas. $\mbox N_2$ minimizes the
multiple, incoherent re-absorption thanks to the quenching of the
resonant fluorescence. The laser radiation  is tuned to excite the
hyperfine transitions between the ground state with total angular
momentum $F_g=3$ and the excited $^2P_{3/2}$ states with $F_e=2,
3, 4$.

The laser is a single-mode edge-emitting pigtail laser ($\lambda$=
852~nm) with 100~mW of laser power and an intrinsic line-width of
less than 5~MHz. Optical feedback is avoided by means of 40~dB
optical isolator and the laser light is coupled into a 10~m long single-mode
polarization-maintaining fiber. The laser head
containing the laser chip, the optical isolator and the fiber
collimator is closed in a butterfly housing of only $\mbox
40~\textrm{cm}^3$. The beam coming from the fiber is collimated
and its polarization is transformed from linear to circular using
a quarter-wave plate. In order to increase the light-atom
interaction time the beam waist is expanded to 4.3~mm and the
laser intensity used for atom excitation is reduced to 
$36~\mbox{$ \rm \mu W/cm^2$}$ using a set of neutral filters. The
transmitted laser light is detected and analyzed.

CPT resonance is created when two different
Zeeman sublevels are coherently coupled to a third common Zeeman
sublevel. The three-level system involved is called $\Lambda$
system in the case the third common coupled level is the  highest
in energy, $V$ system in the case it is lowest. In our case a
number of $\Lambda$ and $V$ systems are created with circularly
polarized light preserving the selection rules \mbox{$\Delta m_f =
0,+1$} and \mbox{$\Delta m_f=0,-1$}. Chains of $\Lambda$ systems
are formed on the  $F_g=3 \rightarrow F_e=2,3$ group of
transitions, while chains of V systems are formed on the $F_g=3
\rightarrow F_e=4$.

\begin{figure}[h]
  \centering
  \includegraphics[width=7cm]{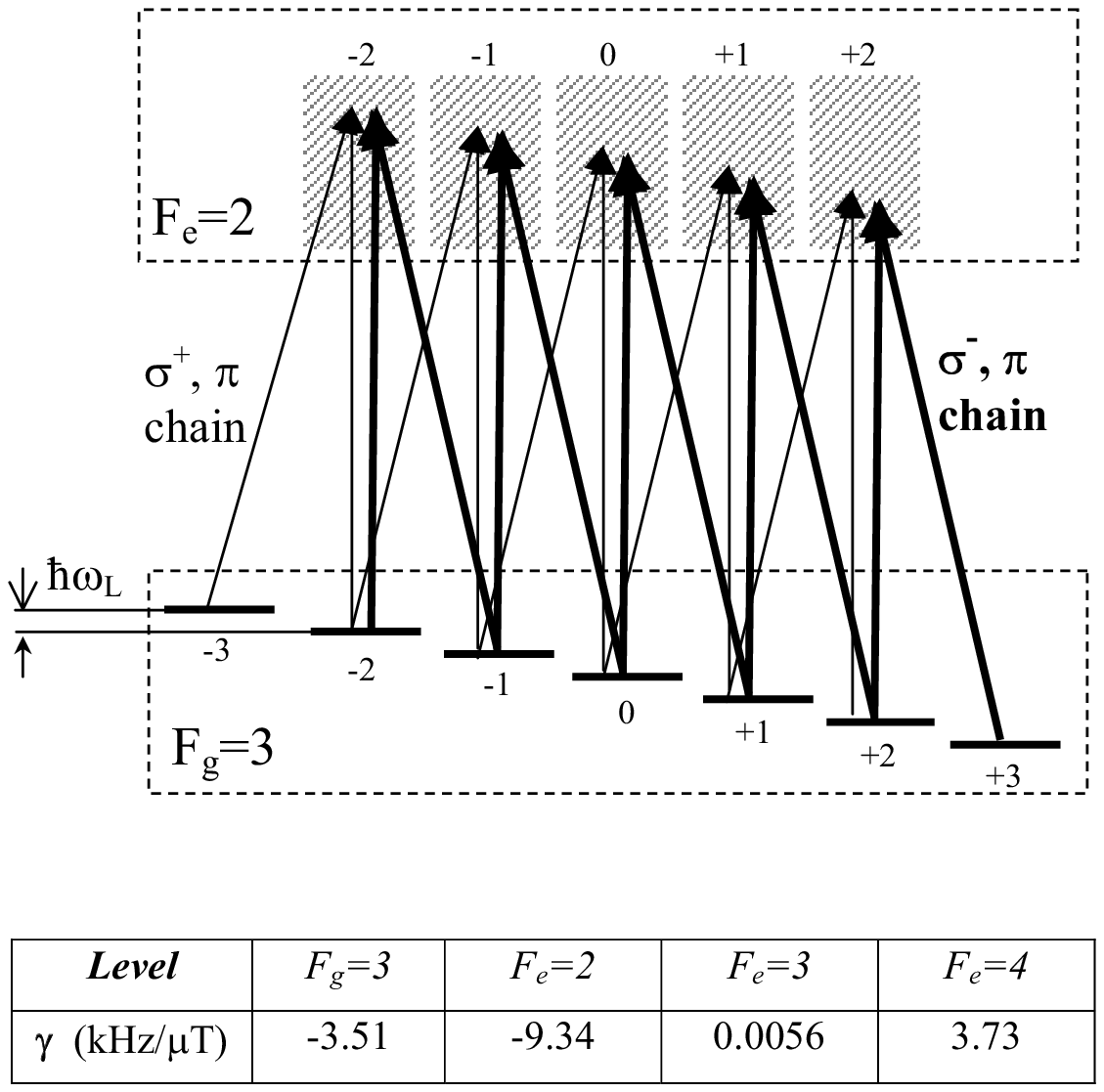}
    \caption{Representation of $\Lambda$-system chains  for the $F_g=3
      \rightarrow F_e=2$ system. The quantization axis is parallel to
the magnetic field and  perpendicular to the laser
beam. Circular polarization is decomposed in two in-quadrature
linearly polarized waves, one of which is in turn decomposed in
two counter rotating fields circularly polarized around the quantization
axis. The complete  scheme would involve also the hyperfine components
$F_g=3 \rightarrow F_e=3,4$. In the table are reported the
gyromagnetic factors $\gamma$ for all the hyperfine levels of
interest. \label{levels}} 
\end{figure}

The magnetic field under measurement breaks the  Zeeman degeneracy
and makes adjacent ($|\Delta m_F|=1$) sublevels  separated by
 $\hbar \mu_0 g_F B = \hbar \omega_L$, where $\hbar$ is the Planck
constant, $\mu_0$ is the Bohr magneton, $g_F$ is the Land\'e
factor of the considered ground-state, $\omega_L$ is the Larmor
frequency and $B$ is the magnetic field strength. In
Fig.\ref{levels}, as example, is represented the schematics of
$\Lambda$-system chains formation for the $F_g=3 \rightarrow
F_e=2$ system of transitions.

The RF signal used to produce couples of suitably separated
frequency components in the spectral profile of the laser emission
is swept in a small interval around the resonant value $\omega_L$.
A set of data acquired along the frequency sweep allows for
visualizing the CPT resonance profile.

\begin{figure}[htbp]
  \centering
  \includegraphics[width=7cm]{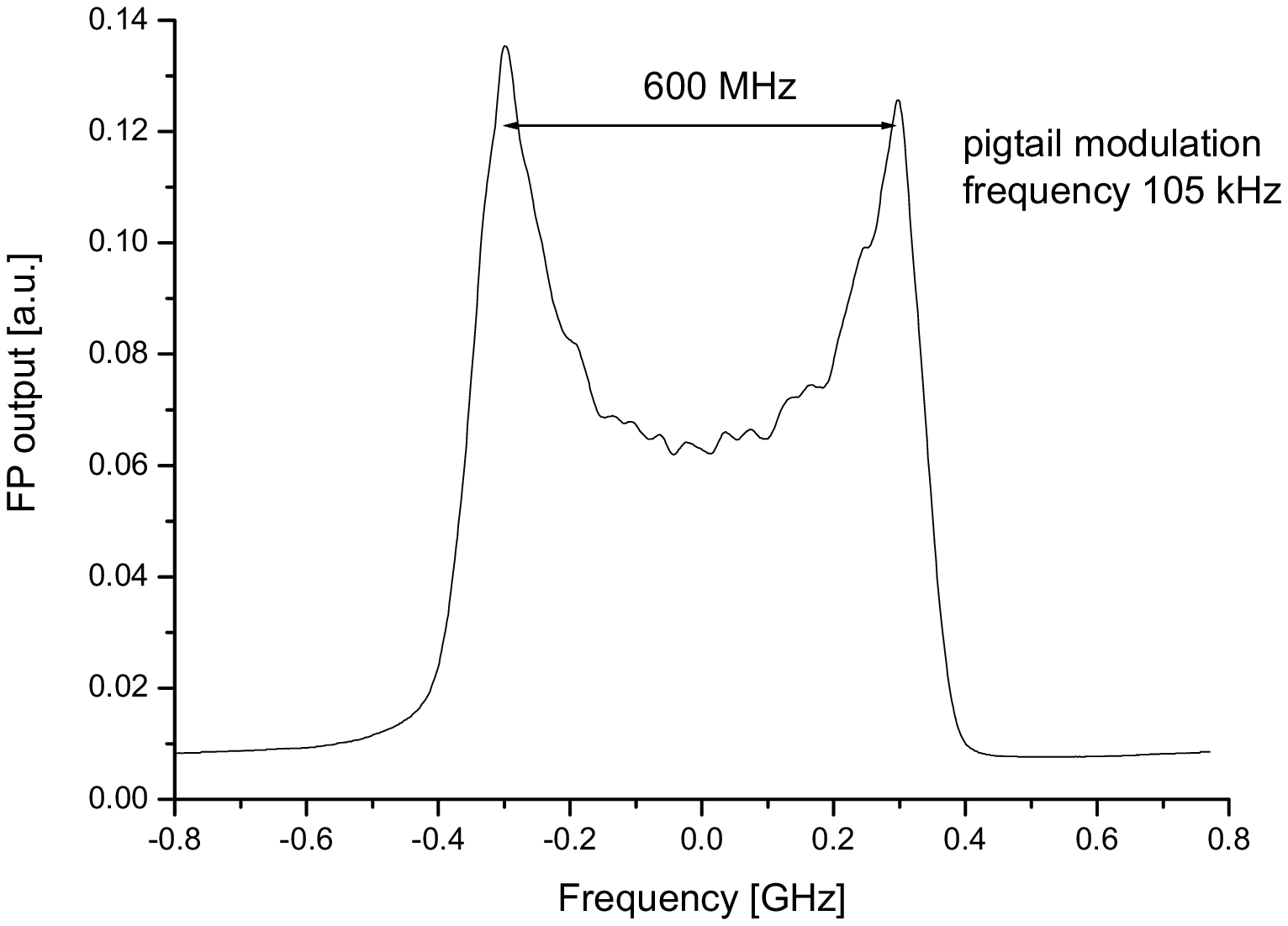}
\caption{Pigtail laser spectrum recorded using a confocal
  Fabry-Perot interferometer with FSR~=~1.5~GHz. The modulation frequency
  is 105~kHz. The inferred modulation index is \mbox{$\sim$
  2800}.\label{pigtail_sp}.}
\end{figure}

As the magnetic field strengths of interest range from few $\mu$T to
few mT, the modulation frequency $\Omega_{RF}$ ranges from few tens of
kHz to few MHz. The $\Omega_{RF}$ is generated by a waveform
generator (Agilent 33250A 80-MHz Function Generator) that is coupled
to the laser  by means of a passive circuit specially designed to make
the response of the laser rather flat in the frequency range of
interest. The circuit uses capacitors to ac couple the RF to the
laser junction, and resistors of rather large value (several hundreds
Ohm) to convert the voltage to current at the output of the
generator. A pair of inductive elements, oriented so that
possible spurious magnetic pick-up is canceled, are
connected in series with the dc supply in order to
prevent the modulating signal from being significantly counteracted by
the laser current driver.  As an overall, this coupling allows for
achieving very large modulation indexes. The envelope of the
unresolved frequency components of the resulting comb of laser
frequencies can be observed using a Fabry-Perot spectrometer (see
Fig.\ref{pigtail_sp}).  

\begin{figure}[h]
  \centering
  \includegraphics[width=7cm]{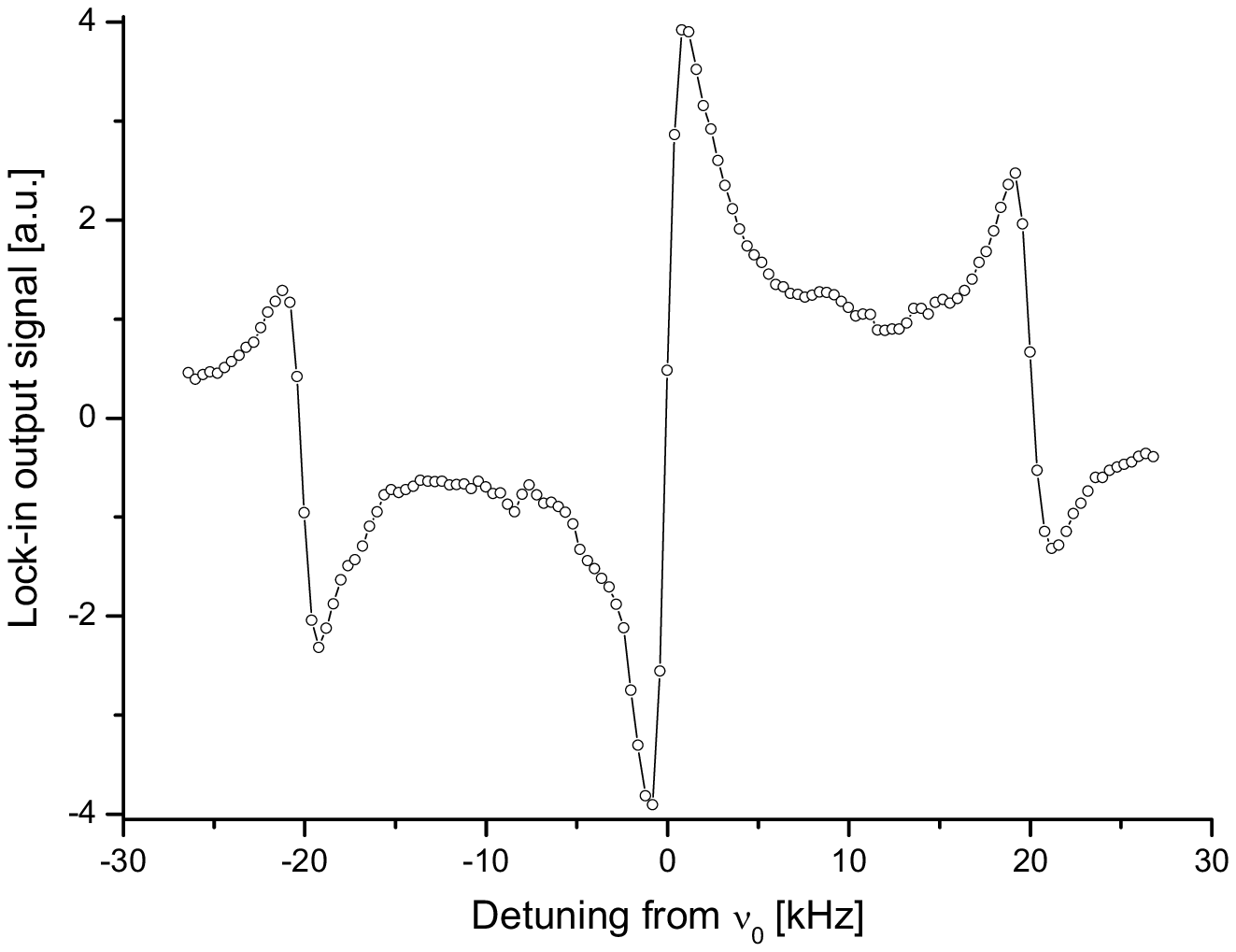}
    \caption{FM spectroscopy of the CPT resonance at modulation frequency of
    20~kHz with deviation of 20~kHz. \label{FM_sp}.}
\end{figure}

In order to increase the S/N of the detected transmitted light, a
Phase Sensitive Detection (PSD) is performed. The $\Omega_{RF}$ is
thus frequency-modulated at $\Omega_{PSD}$ and the atomic
response in phase with a reference signal at $\Omega_{PSD}$ is
extracted. Laser electric field can be expressed by:
\begin{eqnarray}
\vec{E}&=&E_0~(\hat{x}+i\hat{y})~\exp\{i[\omega_0t+\varphi(t)+
\\\nonumber&+&
M_{RF}\cos(\Omega_{RF}t+M_{PSD}\cos(\Omega_{PSD}t))]\}+c.c.
\end{eqnarray}
where $\hat{x}$ and $\hat{y}$ are the usual unitary vectors
perpendicular to the laser wave-vector, $\omega_0$ 
is the optical frequency, $\varphi(t)$ accounts
for the laser linewidth invoking for instance the celebrated
phase-diffusion model, $\mbox M_{RF}$ is the modulation index
of the RF modulation and $\Omega_{PSD}$ is the PSD modulation
frequency with its modulation index $\mbox M_{PSD}$. One typical FM
spectrum of the CPT profile is presented in  Fig.\ref{FM_sp} where
$\Omega_{PSD}=20~\textrm{kHz}$ and $\mbox M_{PSD}=1$. The central
feature is used for magnetic field determination.

An improvement in the noise rejection and hence in the sensitivity
can be achieved using a differential  sensor. Such arrangement 
is very appropriate when a registration of very weak magnetic fields
is desired  \cite{Aff02}. In this case, two identical sensors are
assembled in parallel, at a distance of 11~cm. The light is coupled
to the second arm of the sensor using an intensity beam splitter (see
Fig.\ref{setup}). When evaluating the efficiency of the differential
setup in rejecting the noise, we distinguish three kinds of noise
contribution. The first is due to the detection and
amplification stages. This contribution, which is generally smaller
than the others is increased (nominally by a factor $\sqrt2$). The
second kind is due to intensity fluctuations of the laser emission
(rather small in our case) and to frequency noise of its optical
frequency, in turn discriminated by the Doppler profile. This kind of
noise is effectively rejected. As a third kind one can consider the noise
due to magnetic field fluctuations  generated by magnetic sources different
from the one under examination (e.g. consider  the case of Earth
magnetic field fluctuation while measuring weak biological fields
generated by close sources). In this case, provided that both the Cs
cells are in conditions of CPT resonance, the differential sensor
responds to the (usually very small) gradients of the field generated
by far-located sources (while their common mode field is canceled) and
to the field generated by sources (if any) located very close to one
of the two cells. In this sense depending on the application, the
differential setup can be used either as a gradiometer or as detector
of field variations produced by close sources placed in magnetically
polluted environment. 

The components of both arms of the differential sensor, {\it i.e.}
the fiber collimator, quarter-wave plates, beam expanders, neutral
filters, Cs-N$_2$ cells and the PDs (we use large area, low noise, 
non-magnetic photo diodes) including a reference
Cs-vacuum cell are assembled in a separate plate and can be placed
away from the instrumentation. All used materials are highly
non-magnetic so that  the sensor does not perturb the magnetic
field to be measured. In the condition of our Laboratory an 
improvement of the signal to noise ratio (S/N) of a factor of 5 was
obtained.  

\subsection {\normalsize CPT profile and noise} 
\label{CPT profile and noise}
\begin{figure}[h]
  \centering
  \includegraphics[width=7cm]{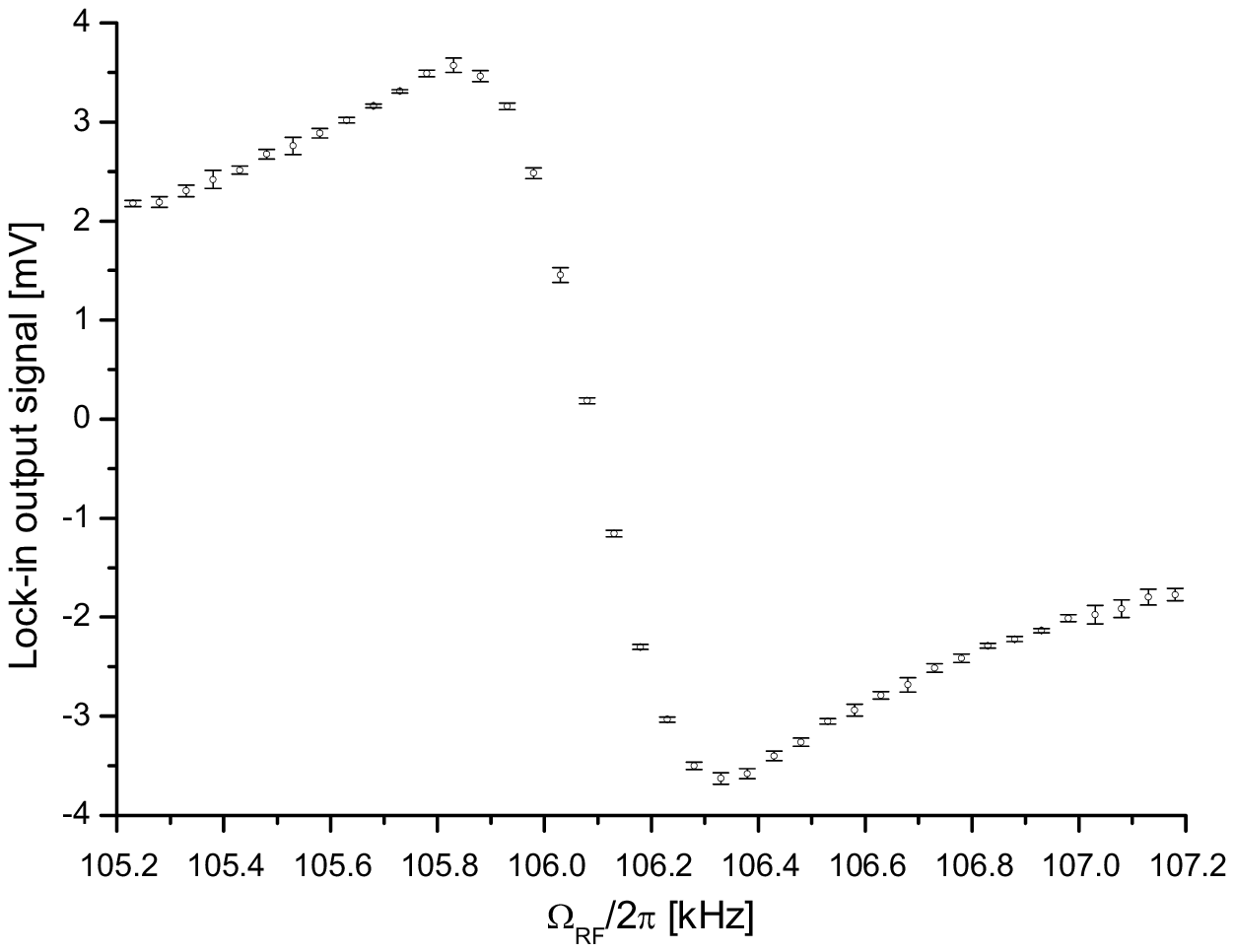}
    \caption{Typical CPT profile observed when scanning the modulation
      frequency around $\omega_L/2\pi$. The lock-in time constant is 30~ms
      with 12~dB/oct output filter which determines the detection bandwidth of
      4.16~Hz. The linewidth resulting from the fit procedure is
      $\Gamma$~=~700~Hz. \label{CPT_profile}.}
\end{figure}

The magnetic field measurement operatively consists in the determination of
the central frequency $\nu_0$  of the resonance profile (2$\pi \nu_0$
is the estimate of $\omega_L$). The registered CPT profile, read at
the output of the lock-in, reproduces the first derivative of the CPT
(reduced absorption) resonance. One typical CPT profile is presented
in Fig.\ref{CPT_profile}. The error bars are evaluated by the lock-in
amplifier from the standard deviation of the output signal in steady
conditions. Routinely, the noise measurement is done only once - after
setting the lock-in  operation parameters, because it is a time
consuming operation,  and can not be performed at each step.

The central frequency  and the linewidth of the resonance are
estimated by means of a best fit procedure. The fitting function
is the first derivative of a Lorentzian profile with central 
frequency $\nu_0$ and FWHM $\Gamma$. It allows for
achieving an excellent agreement of fit with the data
detected in the experimental conditions described above, provided that
the RF scan is performed in a narrow range around the
$\nu_0$. Discrepancies appear for wider ranges of scan. In this
condition we found that adding a secondary small, odd function with
the same center as the principal derived Lorentzian, removes the
discrepancy in the wings, making the fit result insensitive to the
position of the center of the resonance with respect to the center of
the scan. We used another (smaller and broader) Lorentzian derivative,
to take into account such slower decays in the wings.

The uncertainty on the resonant frequency evaluated by the fit
procedure, converted in magnetic field units, is consistent with
the following estimation:
\begin{equation}
       \frac{\Delta B}{\sqrt{\Delta \nu}}=
       \frac{1}{\gamma} \times
       \frac{n\sqrt{\tau}}{\partial V/\partial \nu},
\label{deltaB_min}
\end{equation}
where $n$ is the noise level at the output of the lock-in, $\tau$
is the measuring time (it depends on both the time-constant and
the slope of the output filter in the lock-in), $\partial
V/\partial \nu$ is the slope of the CPT curve and $\gamma$ is the
atomic gyromagnetic factor.

The noise level $n$ can be estimated, as written above, by direct
measurement from the lock-in amplifier, while the resonance
central slope is simply related to the ratio between the amplitude
and the FWHM of the signal. In typical working conditions, $n$
amounts to about 3 times the photo-current shot-noise level, and is
mainly due to fast laser frequency fluctuations (the measured
noise decreases to the expected shot-noise level provided that the
optical frequency is tuned out of resonance), while the FWHM
line-width is about 700~Hz. The CPT parameters, 1/$\gamma$
factor in our configuration and the noise pattern of our registration
system, including the magnetic noise in the Laboratory, set the ultimate 
sensitivity of the sensor to 260~pT/$\sqrt{\mbox{Hz}}$ according 
to Eq.~\ref{deltaB_min}. This sensitivity limit is far above the very
ultimate theoretical limit of the sensitivity, which considering  only
contribution of the  light-atom interaction volume and Cs-Cs spin
exchange collisional rate is 8~$\textrm{fT}/\sqrt{\mbox{Hz}}$
\cite{All02}.  

In general, in order to improve the sensitivity limit versus the
theoretical one,  one  has to reduce the resonance line-width $\Gamma$,
to increase the signal-to-noise ratio and to work in a highly shielded
room. 

The main CPT resonance broadening mechanism, in our case, is the limited
light-atom interaction time and contributes to the total linewidth by 
an amount of the order of 600~Hz. Unshielded environmental conditions
and, in particular, magnetic field gradients and AC magnetic fields,
also contribute to the CPT line broadening. The maximum magnetic field
gradient is estimated to be in the range of $80~\textrm{nT/cm}$,
leading to, worst case, a broadening contribution of the order of
280~Hz. AC magnetic fields contribute, instead, mainly with the 50~Hz
and its harmonics spectral components,  with an overall intensity of
40~nT, determining a corresponding broadening of about 140~Hz. Smaller
broadening contributions are furthermore given by the light
shift and the non-linear Zeeman effect \cite{And03}.

\section{\normalsize PC automated controls}
The experimental procedure is totally controlled by a dedicated
LabView program. The program is used to synchronously communicate
with the lock-in amplifier and the RF wave generator so that the
CPT response can be recorded in appropriate and reproducible
conditions.

\subsection{\normalsize Automated registration of the CPT profile and
  subsequent fit}
The CPT profile is registered by querying the lock-in output after
having set the RF and having waited the settling time.
Alternatively, as described below, the program performs
numerically PSD, using large data sets produced by a 16 bit ADC
card. The program also contains routines devoted to an on-line
analysis of the collected data. In particular, after that each RF
scan is completed, a minimum $\chi^2$ fit procedure is launched
to determine the parameters of CPT resonance profile.

When performing numerical PSD, in  contrast with what reported in
Section \ref{CPT profile and noise}, the noise level is evaluated at each
$\Omega_{RF}$ step. In both cases, we obtain rather good values
for the minimized $\chi^2$/DoF (degrees of freedom), demonstrating 
the reliability of the noise estimation and the suitability of the 
fitting procedure.

\subsection {\normalsize Numerical PSD}
The external lock-in amplifier (bulky and expensive) can be
replaced by a compact ADC card. We successfully tested and used a
system based on a commercial 16 bit,  50~kS/s card USB interfaced
to the PC and sets to operate at 40~kS/s. The principle of
operation was slightly changed with respect to the one of the
lock-in amplifier. Specifically, the RF is externally frequency
modulated at 20~kHz using a square wave signal obtained by scaling
by two the  frequency of a clock signal generated by the ADC card.
Consequently, the ADC  data array $y$ corresponds to high and low
values of the RF, accumulated in the even ($y_{2i}$) and odd
($y_{2i+1}$) elements, respectively. The PSD signal and its
uncertainty are then obtained by considering the $N$-size array of
differences $\delta_i=y_{2i+1}-y_{2i}$ and evaluating the average
$\langle \delta \rangle= \Sigma_i \delta_i/ N$ and the standard
deviation $[ \Sigma_i (\delta_i -  \langle \delta
\rangle )^2/ N]^{1/2}$ scaled by   $\sqrt {N}$, respectively.

The number $2N$ of the acquired data is chosen accordingly to the
desired integration time of the  PSD system, and its upper limit
is set by the size of the data buffer in the ADC card
(64~kB~/~2~Bytes~=~32767 readings which corresponds to 0.8~sec
integration time). It is worth noting that the relatively large
amount of data to be transferred makes the choice of USB 1.0
devices not very advantageous, because it introduces a relevant
dead-time at each measurement.

\subsection{\normalsize Frequency stabilization on the Doppler profile}
Multi-frequency diode laser comb excitation  is suitable for
producing narrow and high contrast CPT signals in free-running
lasing conditions. The optical frequency stabilization by the
laser current and temperature controllers provides the needed
short-term stability. Such passive stabilization method
works well  over time intervals of the order of 1~min, but over
longer time-scales slow drifts of both temperature and current make
it unsuitable. For this reason a long-term active stabilization system
must be employed, allowing for relatively rough (accurate within some MHz)
but reliable re-adjustments of the optical frequency.

We adopted a simple method based on a commercial USB ADC-DAC card
with 12 bit resolution, which periodically and automatically (e.g.
once per minute) performs a scan all over the Doppler profile,
numerically determines center and width of the absorption curve,
and finally provides a dc signal, which, sent to the modulation
input of the laser current driver, establishes detuning with
respect to the maximum of the absorption  profile, in terms of the
measured linewidth. In spite of the low cost and simplicity, such
sub-system was demonstrated to be very effective and reliable for
long-term compensation of the optical frequency drifts.
Furthermore it made the whole system comfortable to be operated
during the optimization as well as suitable for applicative
long-lasting use.

Additionally, we note that such approach, at the expense of
periodically suspending the CPT measurement for a few seconds,
does not need any additional laser modulation (which would have
effect on the CPT measurement) or external modulation elements
(such as Electro optical modulators), which would make the set-up
much more complex and expensive.

We have quantified the root mean square fluctuations $\Delta
\nu_{opt}^{rms}$ of the optical frequency, i.e. the center of the
broadband laser spectrum used for CPT creation, observing the apparent
fluctuation of the fitted Doppler maximum position when scanning the
laser current in the same range of nominal values over the Doppler
absorption. From the observed rms variation of the fitted maximum we get
$\Delta \nu_{opt}^{rms}\simeq2 \rm{MHz}$ over time scales of the order
of 1~sec.

\subsection{\normalsize Servo locking at the center of the CPT resonance}
When a higher rate in the magnetic field measurements is required the
time needed to perform RF scan in order to determine the center of the
CPT resonance can be avoided performing single readings of the lock-in
output. This fast operation uses the central, essentially linear,
slope of the CPT resonance (the Lorentzian derivative profile)
resulting from the fit, to convert the lock-in output voltage in
frequency units and hence in field units. 

Initially, one complete $\Omega_{RF}$ scan is accomplished, the fitting
procedure runs,  and, provided that the $\chi^2$ /DoF is reasonably
close to unity, the best-fit parameters  are passed to the routine
devoted to evaluate the field from single readings of the lock-in  and
to keep $\Omega_{RF}$ locked to the resonance center. 

The fit procedure gives the values of both the slope $\frac{dV}{d\nu}$
and the offset $V_0$ at the center of resonance $\nu_0$.  When the
single reading procedure starts working, $\Omega_{RF}$ is set at
$\nu_0$, and the lock-in output $V$ is queried. The deviation $\Delta
V=V-V_0$ is used in a linear approximation to obtain a new  estimate
of the central frequency $\nu_0+\Delta V / (dV/d\nu)$ and hence of the
field. At each step, $\Omega_{RF}$ is updated to the new estimated
$\nu_0$ in order to  keep the system working at the center of the CPT
resonance, with the double aim of maintaining the linear estimation
appropriate, and preventing possible large drifts of the  field from
bringing the system out of the CPT resonance. 

The lock-in time-constant can be selected with different values for
the scan and the single-reading operations. Obviously, to obtain a
comparable noise rejection in  single-reading operation it is
necessary to increase the time-constant. The lock-in settling time
derived from the time-constant and the lock-in output  filter slope is
taken into account in the lock operation in order to update
$\Omega_{RF}$ with a rate $R$ allowing for locking the system at the
actual center of resonance with the maximum speed, but without risks
of oscillations.  

$\Omega_{RF}$ is actually updated at the rate $R^{\prime}$ of the readings 
(this value is limited by the RS232 communications, and generally
exceeds $R$)  consequently, the $\Omega_{RF}$ increment is scaled by a
factor $R/R^{\prime}$.    

The evaluated magnetic field is immediately saved on disk, possibly
simultaneously  with other reference signals (for instance, in the
perspective of applications in  magneto-cardiography,  ECG signals
will be saved as a reference, in view of  offline analysis to be
performed over long-lasting acquisitions).

\section{\normalsize Signal optimization, performances and limits}
\begin{figure}[h]
  \centering
  \includegraphics[width=7cm]{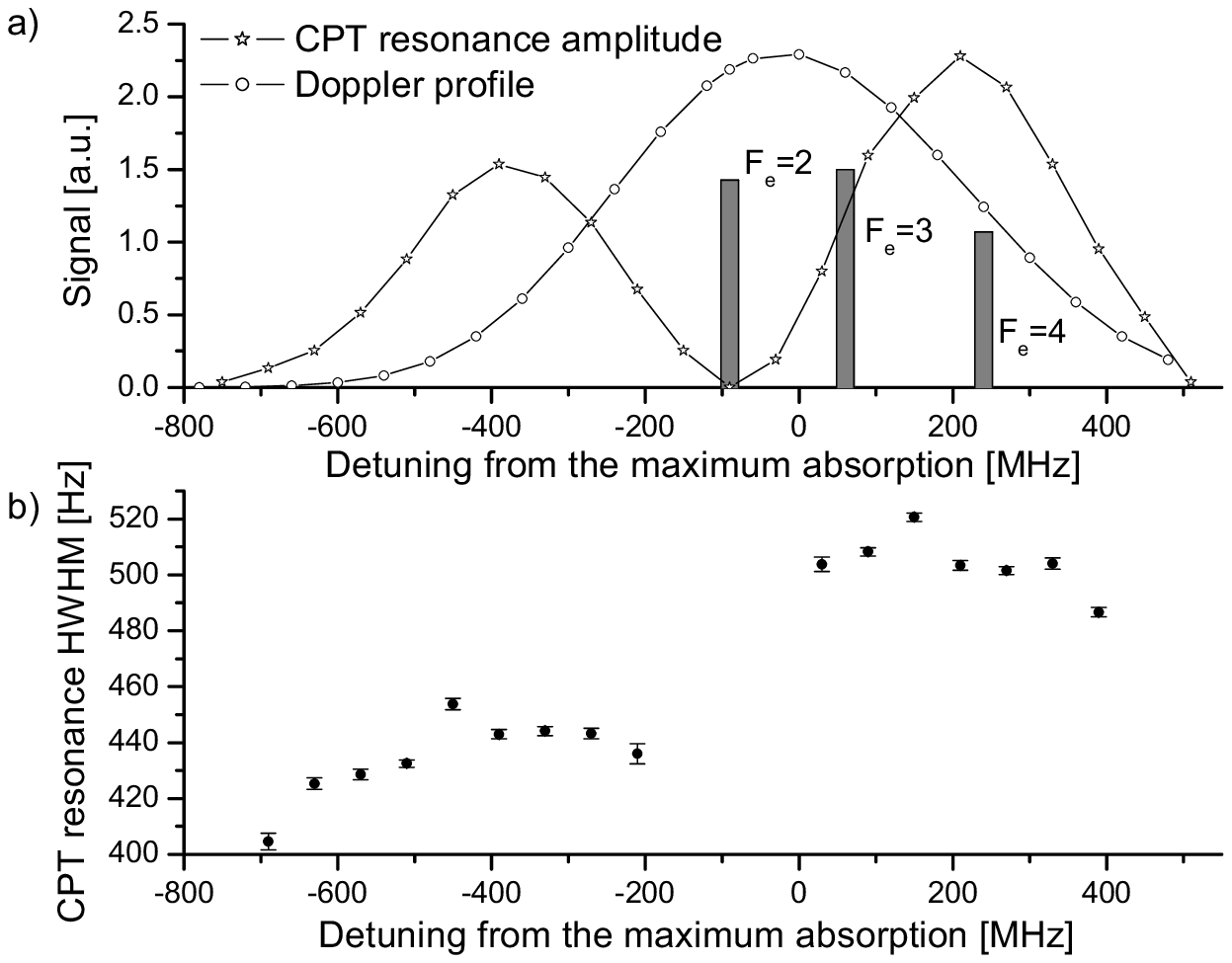}
\caption{a)~CPT resonance amplitude versus optical detuning
(zero frequency corresponds to the maximum of the Doppler
profile). The frequency position of the three hyperfine transitions
are marked with respect to the calculated Doppler  profile. b)~CPT
HWHM versus optical detuning. }\label{CPT_ampl_LS} 
\end{figure}

The dependence of the amplitude of the CPT resonance as a function
of the laser optical detuning  is shown in Fig.\ref{CPT_ampl_LS}a,
where the CPT amplitude, the Doppler broadened fluorescence line and the
frequency position of the hyperfine transitions are reported. The
CPT resonance amplitude shows a two-lobe structure that reflects
the bridge-shaped spectral intensity profile shown in
Fig.\ref{pigtail_sp}, with the two maxima separated nearly by the 
same amount (about 600~MHz). The reason for the vanishing resonance at
intermediate detunings lies in the opposite phase of the beating at
$\Omega_{RF}$ of the FM laser spectrum. Actually, each couple of
adjacent sidebands in the laser spectrum, with their amplitudes
$J_m$, $J_{m+1}$, which produce a beating signal at $\Omega_{RF}$,
contains one odd and one even Bessel function so that, due to the
fact that $J_{-m}(M)=(-1)^m J_m(M)$ the beat phases is opposite
for the couple  (m,~m+1) and (-m,~-(m+1)) respectively, i.e. for
the couples belonging to the right and left wings of the bridge.
From this point of view, depending on the optical detuning, a
synchronous excitation \cite{budk_rev02,Aco06} of different
velocity classes of atoms is performed, having a unique phase in
the case where one side only of the bridge is in resonance with
the Doppler profile, and two opposite phases when the bridge
center coincides (about) with the Doppler center. A deeper
analysis puts in evidence that the two lobes in Fig.\ref{CPT_ampl_LS}a are
different in amplitude, with higher values on the blue side. This
is related to the dominance of the $F_g=3 \to F_e=3$ and $F_g=3 \to
F_e=4$ transitions on the blue wing of the Doppler line. On the
other hand, the CPT resonance vanishes when the laser is tuned in
the vicinity of the  $F_g=3 \to F_e=2$ transition, and the maxima
of the two lobes are symmetric with respect to such detuning,
accordingly with the fact that this latter, closed  transition
gives the most relevant contribution to the CPT.

\begin{figure}[htbp]
  \centering
  \includegraphics[width=7cm]{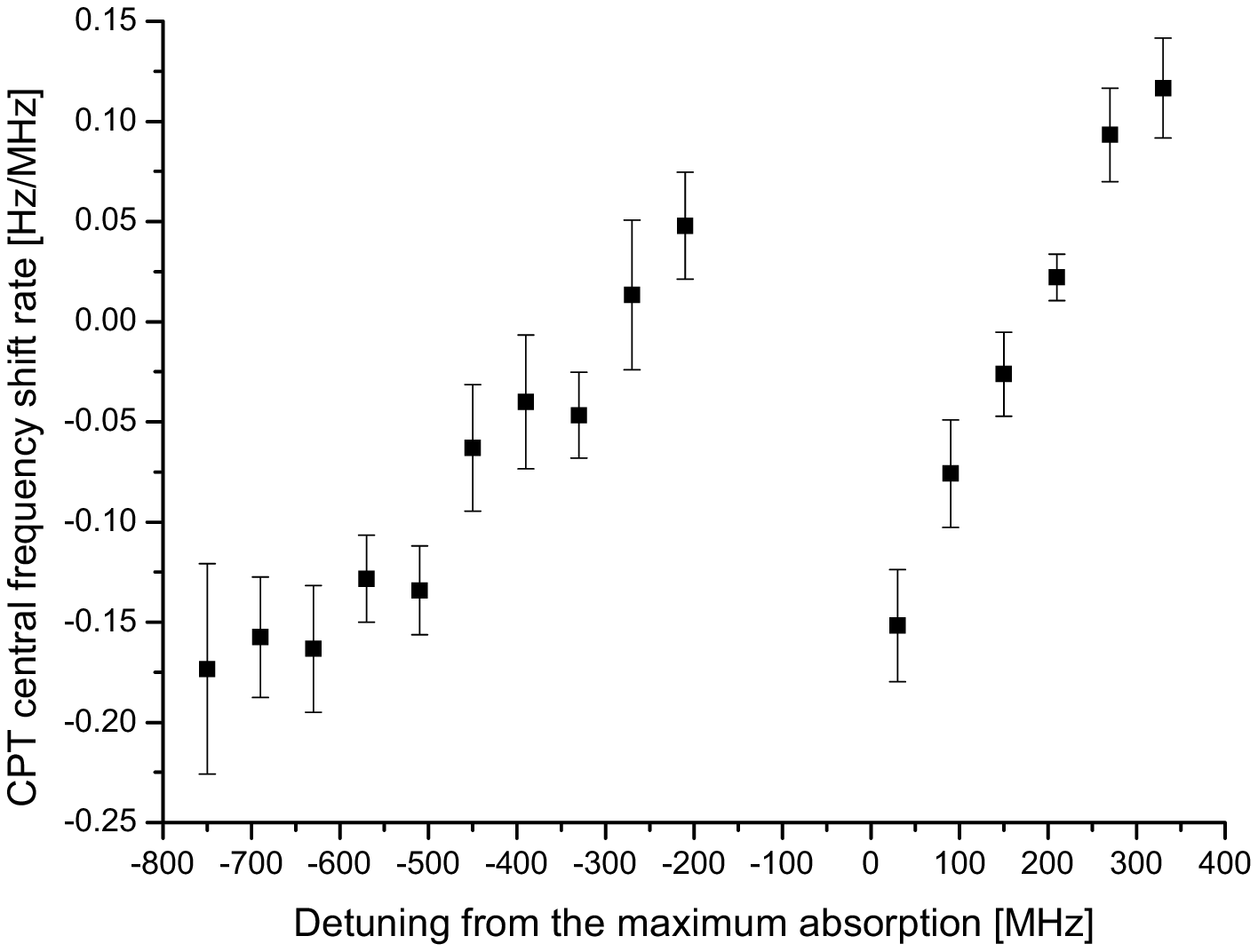}
\caption{CPT resonance center shift rate depending
on optical  detuning for laser intensity of 36~\mbox{$\mu \rm W/cm^2$}. The
value given at  each point is obtained by averaging the
difference in the CPT resonance centers measured 30~MHz above and
30~MHz bellow the corresponding value of the detuning from the maximum
absorption. Error bars represent the standard deviation of each
data set consisting of about 150 measurements.   \label{CPT_LS}}
\end{figure}

Besides the evident variation of the resonance amplitude discussed above,
changing the optical detuning determines also a variation both in the 
resonance linewidth (Fig. \ref{CPT_ampl_LS} b) and in the resonance 
center frequency. The linewidth dependence on optical frequency 
affects mainly the sensitivity of the instrument according with
Eq.~\ref{deltaB_min}. The shift of the CPT resonance center versus 
optical detuning, also called light-shift or ac Stark effect, is due to
the finite dephasing rate among ground states involved in the CPT
preparation (see [\cite{Ari96, Coh92}]). This effect represents an
essential systematic error in the determination of the CPT resonance
center and then affects both the accuracy and the sensitivity of the
magnetometer. The CPT center frequency shift rate versus optical
detuning is presented in Fig.\ref{CPT_LS} for laser intensity of
36~\mbox{\rm $\mu$W/cm$^2$}.

The optimal optical detuning, in view of best magnetometer operation,
is around 400~MHz red detuning, where the light-shift rate passes
through a minimum value and the CPT resonance has lower linewidth.  

\subsection{Magnetic field monitoring}
\begin{figure}[h]
  \centering
  \includegraphics[width=7cm] {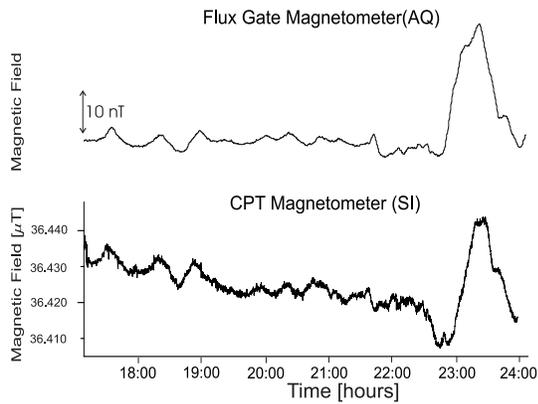}
    \caption{Long lasting monitor of the Earth magnetic field,
    comparison between two different independent measurements. Upper
    trace: acquisition of the fluctuations of the Earth magnetic field 
    modulus measured by the 3-axes flux-gate magnetometer in the
    Geophysical Institute of L'Aquila. Sampling rate
    is one point per minute. Lower trace: acquisition of the modulus
    of the Earth magnetic field in the Physics Department of the
    University of Siena. In this case the sampling rate is about 1
    point each 8~sec. Both the direction and the
    strength of the magnetic field are strongly influenced by the
    presence of ferromagnetic objects in the laboratory and in the
    structure of the building.  \label{long_run}}
\end{figure}

The magnetometer performance was checked by registration of the
Earth magnetic field variation in time. A record of few hours
continuous magnetic field registration with our magnetometer
\cite{noiweb} is shown in Fig.\ref{long_run} together with the
corresponding data of the L'Aquila geomagnetic station
\cite{aquilaweb}.

Both the procedures, when fitting the CPT resonance for determination
of its center and using fast acquisition, were considered and
investigated. In the first one, the whole  CPT profile is registered
and fitted as described above. In this case the measuring time, limited by
the time necessary for CPT profile registration together with the
subsequent data analysis, is of the order of 8~sec. Such procedure
does not make possible to  register fast magnetic field variations
and in this respect can find application in geophysics,
archaeology, material science, etc.

\begin{figure}[htbp]
  \centering
  \includegraphics[width=7cm]{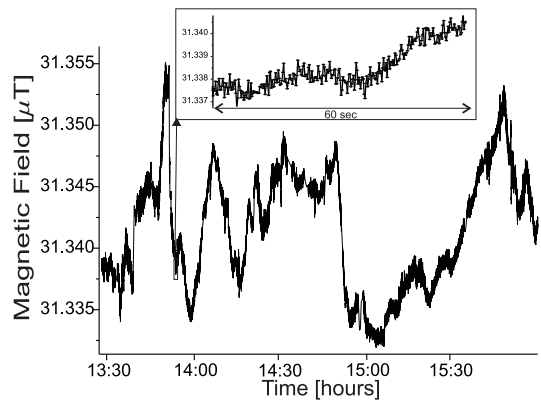}
    \caption{ Magnetic field variation registration in single reading
      operation. In the inset is sketched the zoom over 1~min
      acquisition.\label{lockrun_zoom} } 
\end{figure}

A trace of the Earth magnetic field variation obtained in fast
operation is shown in Fig.\ref{lockrun_zoom}. In this case the
acquisition rate was increased by a  factor of 40 (210~ms per reading). 

\begin{figure}[h]
  \centering
  \includegraphics[width=7cm] {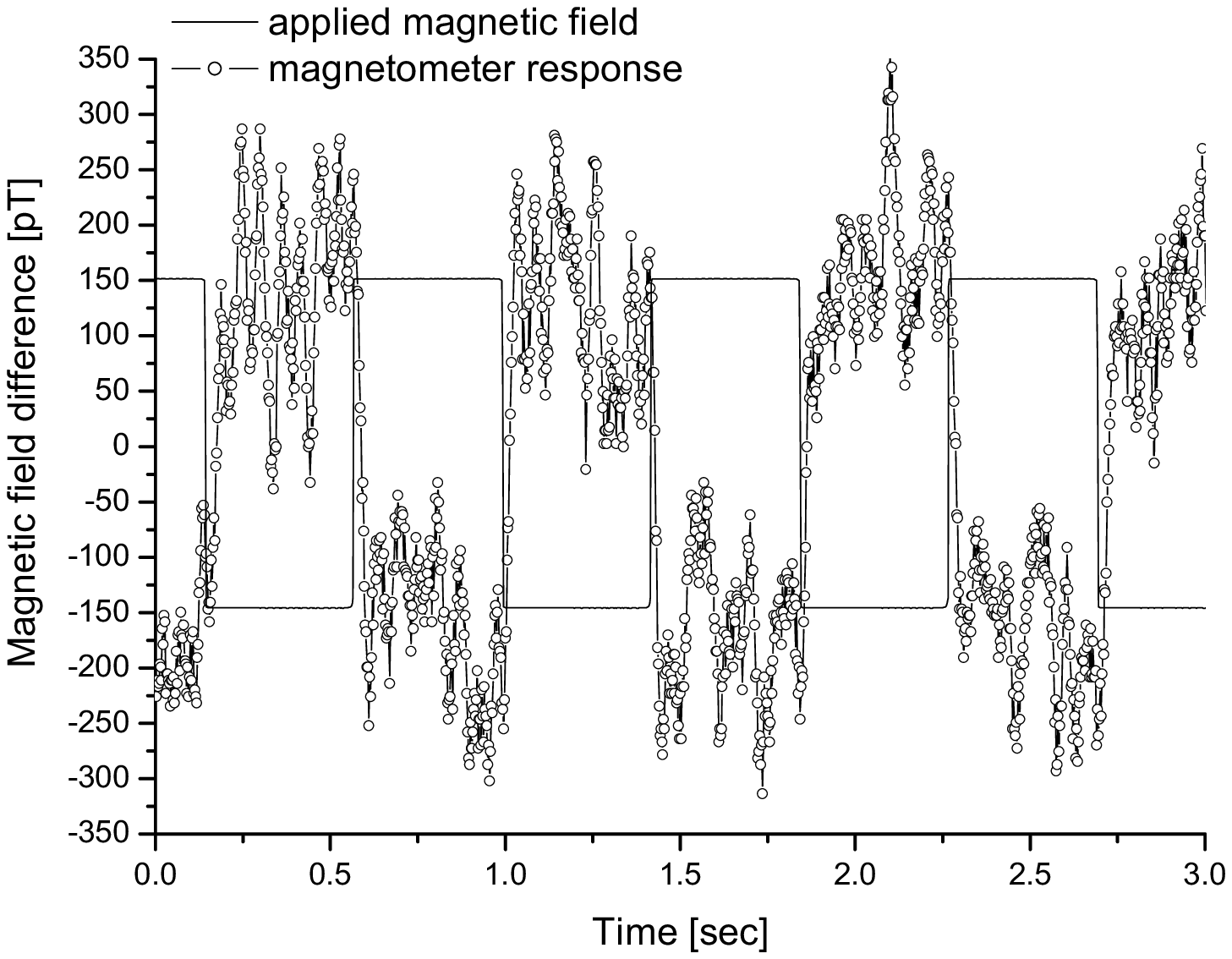}
    \caption{The magnetometer response to
a 300~pT$_{p-p}$, 0.8~Hz square-wave magnetic signal registered
in differential configuration with a band-width determined by the
lock-in time constant of 3~msec, 12~dB/oct output filter and 10
averages. \label{sensitivity}}  
\end{figure}

The magnetometer sensitivity to weak magnetic field variations
superimposed on the Earth magnetic field was determined when working
in the differential configuration. For this purpose, a calibrated
variable magnetic field was applied using multi-turns coil, placed
half a meter away from the sensor. The calibration of the magnetic
field strengths on the two arms of the sensor produced by the coil was
done using the magnetometer itself. The magnetometer response to a
slow and weak variation of the magnetic field in time is presented in
Fig.\ref{sensitivity}. It can be seen that variations in the magnetic
field difference of the order of 300~$\mbox{pT}_{\mbox{p-p}}$ are
well resolved. The inferred magnetometer sensitivity in differential
configuration is  $45~\mbox{pT}/\sqrt{\mbox {Hz}}$. 

\section{Conclusions}
We have built an all-optical magnetometric sensor  supplied  with
a PC-automated  control of the experimental parameters and
an absolute magnetic field measurement data acquisition system.

CPT resonance creation by a kHz-range frequency-modulation of a
free running diode laser in totally unshielded environmental
conditions is demonstrated and an accurate characterization of the
optimal experimental parameters relative to the generation  of
broad band frequency comb spectrum and to the detection strategy
is also given. We presented, furthermore, a detailed analysis of
the dependence of CPT resonance amplitude and width on the optical
frequency tuning, thus determining  the optimal  detuning from the
central frequency of the single-photon absorption spectral
profile. The magnetometer  performs long term continuous
monitoring of magnetic field in the Earth-field range providing a very
sensitive tool for small magnetic field variation registration. We plan to
routinely and systematically publish such data almost in real-time
(preliminary sets are available in Ref.[\cite{noiweb}]), with the aim of
making our system useful for remote Earth magnetic field continuous
observations and possible comparisons with measurements performed
elsewhere.

The best sensitivity, inside a totally unshielded environment,
reached in the differential balanced configuration is $45~\mbox
{pT}/\sqrt{\mbox {Hz}}$.

\section*{Acknowledgments}
We thank S.\ Cartaleva and A.\ Vicino for useful discussions and A.\
Barbini for the technical support. This work was supported by the
Monte dei Paschi di Siena Foundation.

\end{document}